\documentclass[showpacs,twocolumn,prl]{revtex4}

\usepackage{amssymb}
\usepackage{amsmath}
\usepackage{epsfig}

\newcommand{\Df}{\Delta F}
\newcommand{\DfPAP}{\Delta F_{P,AP}}
\newcommand{\Hrep}{{\cal H}_{\mathrm{rep}}}

\newcommand{\ab}{_{\alpha\beta}}

\begin{document}

\title{Interface energies in Ising spin glasses}
\author{T. Aspelmeier}
\affiliation{Department of Physics and Astronomy, University of
Manchester, Manchester M13 9PL, UK}
\author{M. A. Moore}
\affiliation{Department of Physics and Astronomy, University of
Manchester, Manchester M13 9PL, UK}
\author{A. P. Young}
%\email{peter@bartok.ucsc.edu}
\affiliation{Department of Physics,
University of California, Santa Cruz, CA 95064}
\affiliation{
Department of Theoretical Physics, 1, Keble Road, Oxford OX1 3NP,
England}
\date{\today}

\begin{abstract}
The replica method has been used to calculate the interface free energy
associated with the change from periodic to anti-periodic boundary conditions
in finite-dimensional spin glasses.  At mean-field level the interface free
energy vanishes but after allowing for fluctuation effects, a non-zero
interface free energy is obtained which is significantly different from
numerical expectations.
\end{abstract}

\maketitle

A central concept in the droplet picture of spin glasses is the interface free
energy \cite{BanavarCieplak82,McMillan84a,BrayMoore86,FisherHuse86,%
FisherHuse88b} $\delta F$, and the associated stiffness exponent $\theta$
defined by $\delta F \sim l^\theta$ where $l$ is the length scale of the
excitation.
If $\theta>0$
the spin glass state is stable at finite temperature, whereas if $\theta<0$
at $T=0$ large scale excitations cost very little energy
so the spin glass state will be unstable at finite temperature.
The value of $\theta$ at $T=0$ has
been estimated \textit{numerically}, in many calculations, for short range
spin glass models from the effects of changes in boundary conditions, see
e.g.\ Refs.~\cite{McMillan84b,Hartmann99a,Hartmann99b,HartmannYoung01}.  It is
therefore surprising that no attempt has so far been made to determine the
interface free energy from boundary condition changes using the alternative
``Replica Symmetry Breaking'' (RSB) \cite{Parisi79,MezardEtAl87} scenario for
the spin glass state. In this paper we calculate $\theta$
\textit{analytically}
in high dimensions
using the replica method
and show that it conflicts with expectations from
the droplet picture, and numerical work on systems in lower dimensions.

We define the interface free energy in the standard way as the root mean
square change in the free energy of a spin glass when the boundary conditions
along one direction (the $z$ direction) are changed from periodic to
anti-periodic, i.e. $\delta F=\sqrt{\overline{\DfPAP^2}}$ (here and in the
following, the overbar means averaging over bond configurations) where $\DfPAP
= F_{P} - F_{AP}$, and $F_P$ and $F_{AP}$ are the free energies with periodic
and anti-periodic boundary conditions respectively. Anti-periodic boundary
conditions can be realized by reversing the sign of the bonds crossing a plane
whose normal is parallel to the given direction.  It follows that
$\overline{\DfPAP}= 0$.  We note that earlier attempts to calculate a defect
energy \cite{FranzEtAl94,BrezinDeDominicis02} did not employ a definition of
it which is relevant to the droplet picture or numerical studies.

It is convenient to replicate the system with periodic boundary conditions $n$
times and the system with anti-periodic boundary conditions $m$ times, and
keep $n$ distinct from $m$ for the time being. Expanding the replicated
partition function in powers of $m$ and $n$, and taking the logarithm, we have
\begin{multline}
-\ln \overline{ Z_P^n Z_{AP}^m } = 
(n + m) \, \beta \overline{F}  \\
- {(n+m)^2\over 2}\, \beta^2 \overline{\Df^2}
+ {nm \over 2} \, \beta^2 \overline{\DfPAP^2} +  \cdots ,
\label{lnZnZm}
\end{multline}
where $ \overline{\Df^2} = \overline{F_P^2} - \overline{F_P}^2 =
\overline{F_{AP}^2} - \overline{F_{AP}}^2 $ is the (mean square)
sample-to-sample fluctuation of the free energy, the same for both sets of
boundary conditions $P$ or $AP$, and $\overline{F} = \overline{F_P} =
\overline{F_{AP}}$. Hence, to find the variance of the interface free energy,
$\overline{\DfPAP^2}$, we expand out $\ln \overline{ Z_P^n Z_{AP}^m }$ to
second order in the numbers of replicas, $n$ and $m$, separate out the pieces
involving the \textit{total} number of replicas $n+m$, and take the remaining
piece, which is proportional to $n m$.

Using the standard replica field theory \cite{DeDominicisEtAl98}, we can write
$\overline{ Z_P^n Z_{AP}^m } = \int\mathcal{D}q\,\exp(-\beta\Hrep)$ where
$\Hrep$ is the replica free energy, expressed in terms of the spin glass order
parameter field, $q\ab(x)$. It is given by
\begin{multline}
\beta \Hrep = \int d^d x \, \left[
-{\tau \over 2} \sum_{\alpha,\beta} q\ab^2 +
{1 \over 4} \sum_{\alpha,\beta} (\vec{\nabla} q\ab)^2  \right. \\
\left.
-{w \over 6} \sum_{\alpha,\beta,\gamma}  q\ab q_{\beta\gamma}q_{\gamma\alpha} 
-{y \over 12} \sum_{\alpha,\beta} q\ab^4 
\right] ,
\label{Hrep}
\end{multline}
where $q_{\alpha\beta}$ is a symmetric matrix with $q_{\alpha\alpha}=0$, we
have omitted some unimportant terms of order $q^4$, and set $\tau = 1 -
T/T_c$. The fourth order term included is the one responsible for replica
symmetry breaking. The coefficients $w$ and $y$ are arbitrary positive
parameters. The replica indices go $\alpha,\beta,\gamma=1, 2, \cdots, n,
n+1,\cdots, n+m$. The order parameter $q$ divides naturally into blocks of
size $n$ and $m$. From now on, Greek indices will label the first block, Roman
ones the second block, so, for example, $q_{\alpha i}$, means $\alpha\in[1,n]$
and $i\in[n+1,n+m]$, and refers to the respective entry in the off-diagonal,
or mixed, sector.

We shall assume that there is only spatial variation in the $z$ direction,
which we shall take to be of length $L$.  All directions perpendicular to the
$z$ direction are of length $M$. The volume of the system is $V=M^{d-1}L$.
Along the $z$-direction, we impose the boundary condition that the solution is
periodic in the Greek and Roman sectors, and is antiperiodic in the mixed
sectors reflecting the sign reversal of the plane of bonds in the one sector
with respect to the other:
\begin{equation}
\begin{split}
q_{\alpha\beta}(z) &= 
  q_{\alpha\beta}(z+L)  \\
q_{ij}(z) &=
  q_{ij}(z+L)  \\
q_{\alpha i}(z) &=
  -q_{\alpha i}(z+L).
\end{split}
\label{bcAP}
\end{equation}

At mean-field level, there is the following \textit{stable} solution for $\ln
\overline{Z_P^n Z_{AP}^m}$:
%is given by
\begin{align}
-\ln \overline{Z_P^n Z_{AP}^m} &= \beta\Hrep\{q^{\text{SP}}\},
\end{align}
where 
\begin{align}
q^{\text{SP}} &=
\left(\begin{array}{c|c}
Q^{(n)}\ab & 0 \\ \hline
0  & Q^{(m)}_{ij} 
\end{array}\right)
\end{align} 
is independent of the spatial coordinates and $Q^{(s)}$ is a Parisi symmetry
broken saddle point solution of size $s\times s$, with the necessary
modification for finite positive $s$ as derived in \cite{AspelmeierMoore02},
i.e.
\begin{align}
\label{Qdef}
Q^{(s)} &= \lim_{p\to\infty}
\underbrace{\left(\begin{array}{ccc}\cline{1-1}
\multicolumn{1}{|c|}{P^{(s/p)}}&& 0\\ \cline{1-1}
&\ddots& \\ \cline{3-3}
0&&\multicolumn{1}{|c|}{P^{(s/p)}} \\ \cline{3-3}
\end{array}\right)}_{\text{$p$ blocks}},
\end{align}
where $P^{(s/p)}$ is a `standard' Parisi matrix. The limit $p\to\infty$ in
Eq.~\eqref{Qdef} should be interpreted in the same sense as for a standard
replica symmetry breaking procedure, i.e.\ as taking $p$ to infinity when it
is convenient during a calculation.

It is natural that the diagonal blocks are the same as the regular Parisi
ansatz because ordering in the system with periodic boundary conditions, say,
should not be affected by there being another \textit{completely independent}
copy with different boundary conditions. 
Choosing the mixed sector to vanish
seems to be consistent with the standard
interpretation~\cite{MarinariEtAl00} of RSB in
short-range systems, namely that changing the boundary conditions changes the
system \textit{everywhere}. More precisely the surface of the domain wall
separating the regions which flip from the regions which don't flip is space
filling. In this situation, one can reasonably expect zero overlap between
configurations with different boundary conditions. 
%The mixed sector $q^{\text{SP}}_{\alpha i}$ vanishes since 0 is the only
%spatially constant function that is compatible with antiperiodic boundary
%conditions and, for symmetry reasons (the interface may appear anywhere in the
%system), it cannot be a spatially dependent function.

This solution is \textit{identical} to the solution one obtains using the
correct way of breaking the symmetry, as presented in
\cite{AspelmeierMoore02}, for a $n+m$-times replicated system ($n+m$ being
finite) \textit{without} boundary condition changes. We can therefore
immediately use the result from \cite{AspelmeierMoore02} that on mean-field
level, there is no term of order $(n+m)^2$, let alone of order $nm$, and thus
the interface energy vanishes to this order.

We now turn to the loop expansion about the saddle point, which is expected to
be valid for dimension $d$ greater than 6. The first correction is due to
Gaussian fluctuations around the saddle point solution. They are given by
\begin{align}
-\ln \overline{Z_P^n Z_{AP}^m} &= \beta\Hrep\{q^{\text{SP}}\} +
  \frac 12 \sum_k I(k^2),
\end{align}
where
\begin{align}
    \label{Idef}
    I(k^2) &= \sum_{\mu}d_{\mu}\ln(k^2+\lambda_{\mu}),
\end{align}
$k$ is a $d$-dimensional wave vector and $\lambda_{\mu},d_{\mu}$ are the
eigenvalues of the Hessian, evaluated at the saddle point solution, and their
degeneracies. The eigenvalues $\lambda_\mu$ and degeneracies $d_\mu$ are the
same as for a system of size $n+m$ without boundary condition changes (because
the saddle point solution is the same), only the nature of the $k$-vectors
changes for the terms involving eigenvalues whose corresponding eigenvectors
$f$ are nonzero exclusively in the mixed sector (i.e.\ $f\ab=f_{ij}=0$): the
wave vectors have to respect the imposed boundary conditions, which implies
$k=(2n_1\pi/M,\dots,2n_{d-1}\pi/M,(2n_d+1)\pi/L)$ (with $n_i\in\mathbb{Z}$) in
the mixed sector as opposed to $k=(2n_1\pi/M,\dots,2n_{d-1}\pi/M,2n_d\pi/L)$
in the Greek or Roman sectors.

It was shown in \cite{AspelmeierMoore02} for a system without boundary
condition changes that it is initially easier to compute $\partial
I/\partial(k^2)$ than $I$ itself , and that it is given in terms of the
diagonal propagators $G_{\alpha\beta,\alpha\beta}$ (or $G^{xx}_{11}$ in the
limit of infinitely many replica symmetry breaking steps
\cite{DeDominicisEtAl98}) as
\begin{align}
    \frac{\partial I_{P}}{\partial (k^2)} &=
    \sum_{\alpha<\beta}G_{\alpha\beta,\alpha\beta} =
    -\frac n2 \int_{n}^1 dx\,G^{xx}_{n},
\end{align}
where we have dropped the subscript $11$ from the propagators as it is 
irrelevant here and replaced it by $n$ since the propagators depend on it.

Therefore the contribution to $\partial I/\partial(k^2)$ from those
eigenvectors that are nonzero in the Greek or Roman sectors (the periodic
sectors, hence the subscript $P$ below) is
\begin{align}
\label{Igreek}
\frac{\partial I_{P}}{\partial (k^2)} &=
  -\frac n2 \int_{n}^1 dx\,G^{xx}_{n} - \frac m2 \int_{m}^1 dx\,G^{xx}_{m} \\
&=  -\frac{n+m}{2} \int_{0}^1 dx\,G^{xx}_{0} + \frac{n^2+m^2}{2} G^{00}_{0},
\end{align}
The last line follows from the modified symmetry breaking procedure
(Eq.~\eqref{Qdef}), as was shown in \cite{AspelmeierMoore02}. The origin of
the term linear in $n+m$ in Eq.~\eqref{Igreek} is the eigenvectors that are
nonzero in a Parisi block $P^{(n/p)}$ or $P^{(m/p)}$ on the diagonal
\cite{AspelmeierMoore02}, while the origin of the $n^2+m^2$-term is the
eigenvectors that are nonzero in the off-diagonal blocks. This observation
facilitates calculating the contribution from the mixed sector as there are
only eigenvectors of the latter type present, i.e.\ there is no term of linear
order. Therefore $\partial I_{AP}/\partial(k^2)$ is given by
\begin{align}
\label{Imixed}
\frac{\partial I_{AP}}{\partial (k^2)} &= nm G^{00}_{0},
\end{align}
where the prefactor $nm$ reflects the number of eigenvectors in the mixed
sector.

The integral $\int d(k^2)\,G_{0}^{00}$ and the constant of integration have
been worked out in \cite{AspelmeierMoore02}, resulting in
\begin{align}
\begin{split}
    J(k^2) &:= \int d(k^2)G^{00}_{0} 
    = \ln(k^2+\frac{x_{1}^2w^2}{2y}) \\
&\quad - \frac{4w(4yk^2 + wx_{1})}{4yk^2\sqrt{4yk^2+w^2x_{1}^2}}
     \tan^{-1}\frac{wx_{1}}{\sqrt{4yk^2+w^2x_{1}^2}},
\end{split}
\end{align}
where $x_1$ is the breakpoint of the Parisi $q$-function.  We can now assemble
in $I$ the terms of quadratic order,
\begin{align}
\label{Ione}
I &= (n+m)C + \frac{n^2+m^2}{2}J_{P}(k^2) +
nm J_{AP}(k^2) \\
\label{Itwo}
&= (n+m)C + \frac{(n+m)^2}{2}J_{P}(k^2) +
nm(J_{AP}(k^2)-J_{P}(k^2)).
\end{align}
The constant $C$ is of no interest to us. The subscripts $P$ and $AP$ on $J$
mean that $J$ must be taken as $0$ when the argument is not of the required
type, i.e.\ periodic or antiperiodic.

We can now identify the term that gives rise to the interface
energy. Comparison with Eq.~\eqref{lnZnZm} shows
\begin{multline}
\label{ienergy}
\beta^2\overline{\DfPAP^2} = 
  \left({\sum}_{AP}-{\sum}_{P}\right) J(k^2) = \\
\sum_l \sum_{r=-\infty}^{\infty} 
\left(J(l^2+\frac{(2r+1)^2\pi^2}{L^2}) - J(l^2+\frac{(2r)^2\pi^2}{L^2})\right)
\end{multline}
where the subscripts on the sums indicate the nature of the allowed
$k$-vectors, as made explicit in the second part of the equation where the $z$
component of the $k$-vector has been split off, leaving the $d-1$-dimensional
wave vector $l$. The sum over the $z$ component has been extended to
$\pm\infty$, introducing only exponentially small errors for large $L$.

We note a potential pitfall in this result: the contribution to
Eq.~\eqref{ienergy} from the $k=0$ term (in ${\sum}_P$) diverges. Usually,
this problem is removed by converting the sums to integrals converging in high
enough dimensions, and arguing that the divergence is, in reality, only a
subdominant contribution. However, since $\theta<(d-1)/2$ \cite{FisherHuse86},
the interface energy is subdominant itself, so it is not clear whether the
subdominant terms from the $k=0$ mode are in fact dominating over the terms we
kept. Therefore we need to treat the $k=0$ mode properly before proceeding.
The way to do this is to go to the equation of state for $q\ab$ and include
the $k=0$ mode exactly, while treating the other modes perturbatively as
before. The complete equation of state is given by Eq.~(15) from
\cite{DeDominicisEtAl98}, and restricted to the $k=0$ mode it reads
\begin{multline}
2\tau q\ab + w (q^2)\ab + \frac{2y}{3}q\ab^3  = \\
-\frac{1}{V}
\left(w \sum_{\gamma\ne\alpha,\beta}G_{\alpha\gamma,\beta\gamma}(k=0) +
2yq\ab G_{\alpha\beta,\alpha\beta}(k=0)\right) .
\end{multline}
This equation is highly nontrivial since $G$ in this expression is the
\textit{full} propagator. We do not propose to solve this formidable
self-consistency equation, but we note that the presence of the right hand
side shifts $q\ab$ by an amount $\epsilon\ab$ from the mean-field value, which
in turn shifts the eigenvalues of the Hessian. The left hand side is given by
$\sum_{\gamma\delta}G^{-1}_{\alpha\beta,\gamma\delta}(k=0)
\epsilon_{\gamma\delta} = \mathcal{O}(\epsilon)$ (recalling that $G^{-1}(k=0)$
is equal to the Hessian), the right hand side is of order
$1/V\lambda_{\text{min}}$, where $\lambda_{\text{min}}$ is the smallest
eigenvalue of the Hessian. If $\lambda_{\text{min}}=\mathcal{O}(\epsilon)$,
which is the natural expectation, it follows that $\lambda_{\text{min}}\sim
V^{-1/2}$. Therefore $G(k=0)$ has changed from being infinite to being of
order $V^{1/2}$. This argument is not rigorous, however, therefore we prefer
to denote the exponent more generally by $2 \mu$.  The upshot of this treatment
is that we can exclude the divergent $k=0$ terms from the sums over wave
vectors (as they have been dealt with non-perturbatively), provided a term of
order $V^{2 \mu}$, where $\mu$ may be $1/4$, is introduced in the $n^2$ and
$m^2$ terms in Eq.~\eqref{Ione}. This additional term is \textit{identical} to
the free energy fluctuations in the Sherrington-Kirkpatrick model which has
only the $k=0$ mode, and will be denoted by $\Delta f_{\text{SK}}^2
V^{2 \mu}$. This observation allows us to obtain estimates of $\mu$ from
existing numerical work \cite{CabasinoEtAl88,BouchaudEtAl02,Palassini02},
which supports $\mu=1/4$.

Since we are expecting that the changes to the eigenvalues are of order
$V^{-1/2}$, while the changes due to the different boundary conditions are of
order $1/L^2$, our treatment of the non-zero $k$ modes will be satisfactory in
the range of dimensions where the loop expansion applies, i.e.\ $d>6$.

Upon completing the square as in Eq.~\eqref{Itwo} the contribution $\Delta
f_{\text{SK}}^2 V^{2 \mu}$ appears in the $nm$ term, so from Eq.~\eqref{ienergy}
we get
\begin{multline}
\label{ienergy2}
\beta^2\overline{\DfPAP^2} = \\
\sum_{l\ne 0} \sum_{r=-\infty}^{\infty} 
\left(J(l^2+\frac{(2r+1)^2\pi^2}{L^2}) - 
J(l^2+\frac{(2r)^2\pi^2}{L^2})\right) +
\\
2 \sum_{r=1}^\infty\left(J(\frac{(2r-1)^2\pi^2}{L^2}) - 
J(\frac{(2r)^2\pi^2}{L^2})\right) + \Delta f_{\text{SK}}^2 V^{2 \mu}
\end{multline}

The sums over $r$ in Eq.~\eqref{ienergy2} can be calculated exactly, in
principle, and the sum over $l$ can be converted to an integral with a lower
cutoff and carried out, but the result is too long to show here. The important
feature of it is that the leading behaviour as a function of $L$ is determined
by the divergent part of $J$ as $k^2 \to 0$. The other parts of $J$ only give
exponentially small corrections. Since $J(k^2)\approx-\pi w/4yk^2$ for small
$k^2$, it is sufficient to work out the term
\begin{align}
\frac{-w\pi}{4y}
\sum_{r=-\infty}^{\infty}
\left(\frac{1}{l^2+\frac{(2r+1)^2\pi^2}{L^2}} - 
  \frac{1}{l^2+\frac{(2r)^2\pi^2}{L^2}}\right)
&= \frac{w\pi L}{4yl\sinh lL}.
\end{align}
Together with
$\sum_{r=1}^\infty(\frac{1}{(2r-1)^2}-\frac{1}{(2r)^2})=\pi^2/12$ this gives
\begin{align}
\begin{split}
\label{theta0}
\beta^2\overline{\DfPAP^2} &= M^{d-1}\frac{S_{d-1}}{(2\pi)^{d-1}}
\int_{\frac{2\pi}{M}}^\infty dl\,l^{d-2} \frac{w\pi L}{4yl\sinh lL} \\
&\quad-\frac{w\pi}{24y}L^2 + \Delta f_{\text{SK}}^2 V^{2 \mu}
\end{split}\\
\begin{split}
&=L^2 f^2(L/M) -
\frac{w\pi}{24y}L^2 + \Delta f_{\text{SK}}^2 V^{2 \mu},
\end{split}
\label{theta}
\end{align}
where 
\begin{align}
f^2(L/M) &= \frac{w\pi S_{d-1}}{4y(2\pi)^{d-1}}\left(\frac ML \right)^{d-1}
\int_{\frac{2\pi L}{M}}^\infty\frac{dx\,x^{d-3}}{\sinh x}
\end{align}
is an exponentially decreasing scaling function and $S_{d}$ is the surface of
a $d$-dimensional unit sphere.

Only the first term in Eq.~\eqref{theta} has a form compatible with aspect
ratio scaling \cite{CarterEtAl02}, according to which
$\sqrt{\overline{\DfPAP^2}}=L^{\theta}f(L/M)$.
On the face of it, this would give rise to $\theta=1$ for all
dimensions. The other two terms, however, do not have aspect ratio scaling
form. In particular, the term $\Delta f_{\text{SK}}^2 V^{2 \mu}$, which is
dominant in $d>6$ if $\mu>1/6$, depends only on volume but not on shape.

Our calculation is exact in high dimensions within
the replica symmetry breaking scenario for spin glasses. It is quite unusual
and
contradicts all expectations one might have about the interface
energy based on experience from other systems and numerical data. It is
significantly different from that found in, for instance, ferromagnets. There,
the defect energy comes from the gradient term in the analogue of
Eq.~(\ref{Hrep}).
%and it is finite even in the mean field limit.
Here, on the
other hand, the mean-field solution is independent of $z$ so there is no
contribution from the gradient term.
%and the defect energy vanishes in the mean field limit. It is of order the
%inverse of the lattice coordination
%number $2d$.
A difference between the interface energy in spin glasses and
ferromagnets is, however, that in ferromagnets there is a `real' domain wall,
whereas in spin glasses, the interface can only be defined by comparing one
system to a reference system with the opposite set of boundary conditions.
Thus strictly speaking, the interface in spin glasses is not a physical system
itself, which may account for the absence of an interface energy on the mean
field level.

The failure of aspect ratio scaling is a strong prediction which contradicts
numerical evidence for $d=2$ \cite{CarterEtAl02,HartmannEtAl02}. The replica
symmetry breaking scenario predicts space-filling domain walls
\cite{MarinariEtAl00,PalassiniYoung00}, therefore the dependence of the
interface energy on volume but not on shape (to leading order) appears natural
since the interface explores even the remote corners of the sample and would
be likely to exist in some form down to three dimensions, even though the 
loop expansion used in this paper will need modification below six dimensions.
%The work
%reported here thus
This suggests a simple test of replica symmetry breaking ideas.
If they are valid in three dimensions, then aspect ratio scaling will fail. To
date, there is (weak) evidence that aspect ratio scaling \textit{works} in
three dimensions \cite{CarterEtAl02}.

\begin{acknowledgments}
TA acknowledges support by the German Academic Exchange Service (DAAD). APY
acknowledges support from the National Science Foundation under grant DMR
0086287 and the EPSRC under grant GR/R37869/01. He also thanks David
Sherrington for hospitality during his stay at Oxford. We would like to thank
C. De Dominicis for discussions.
\end{acknowledgments}

\bibliography{Spinglass} 

\end{document}